\documentclass[referee]{raa}            

\usepackage{graphicx,times}             
\usepackage{natbib}
\usepackage{amssymb,amsmath}
\bibpunct{(}{)}{;}{a}{}{,}

\usepackage[pagebackref=true]{hyperref}

\begin{document}

  \title{Quantifying the Relationship Between Galaxy Specific Star Formation Rate And Halo Spin For Star-forming Galaxies}

   \volnopage{Vol.0 (20xx) No.0, 000--000}      
   \setcounter{page}{1}          

   \author{Wenxiao Xue
   \inst{1,2}
   \and
   Zichen Hua
      \inst{1,2}
    \and Yu Rong
      \inst{1,2,*}
   }

   \institute{Department of Astronomy, University of Science and Technology of China, Hefei, Anhui 230026, China; {\it rongyua@ustc.edu.cn}\\
        \and
             School of Astronomy and Space Sciences, University of Science and Technology of China, Hefei 230026, Anhui, China\\
\vs\no
   {\small * corresponding author}}

\abstract{ Utilizing ALFALFA HI data, we investigate the relationship between specific star formation rate (sSFR) and halo spin across various star-forming galaxies. Our analysis reveals weak yet statistically significant positive correlation between sSFR and halo spin, irrespective of the galactic environment. This trend suggests that galaxies with higher spin parameters tend to host dynamically colder, gas-rich disks, sustaining elevated gas surface densities and prolonged star formation. These findings align with theoretical expectations of angular momentum-regulated gas accretion but highlight discrepancies with cosmological simulations, underscoring unresolved challenges in modeling baryonic feedback and star formation efficiency.
\keywords{galaxies: formation --- galaxies: evolution --- methods: statistical}
}

   \authorrunning{Xue, Hua \& Rong }            
   \titlerunning{sSFR weakly depends on halo spin}  

   \maketitle

%
%
\section{Introduction}           
\label{sec:1}

In the standard galaxy formation model, halo spin is considered pivotal in galaxy formation and evolution, influencing morphology and regulating baryonic fraction \citep{Mo98,Guo11,Rong17a,Amorisco16,Benavides23,vandenBosch98,Diemand05,Desmond17}. However, hydrodynamical simulations (e.g., \citealt{Kim13,Jiang192,Yang23,DiCintio19}) have sparked debate on the role of halo spin in low-mass galaxies. For most dwarf galaxies, stellar distributions may be independent of or weakly dependent on halo spin, with concentrations and baryonic feedback exerting more significant influence on low-mass galaxy properties. Halos with shallower potential wells and lower concentrations exhibit stronger feedback effects, leading to reduced star formation efficiencies, lower stellar mass fractions, and more extended stellar distributions \citep{Kravtsov18,Sales22,Hopkins12,Sawala15,Rong25,Liu25}. Despite extensive research, the influence of halo spin on galaxy structure and evolution remains incompletely understood and lacks consensus.

The star formation rate, a crucial parameter defining galaxy evolutionary state, is influenced by internal factors such as feedback, fuel supply, metallicity, and dust content \citep{Springel00,Hjorth14,Grudic18,Hayward11}. Environmental factors also play a significant role in shaping the star formation rate \citep{Gonzalez09,Tinker17}. Among the myriad of galaxy properties, the impact of halo properties on the star formation rate, while not direct, is paramount. Halos' mass and surface density have been shown to significantly affect star formation rates in galaxies \citep{Dahlem06,Kimm09}, yet the influence of halo spin on star formation rate remains uncertain. \cite{Rong24a} propose a link between spin and star formation rate, suggesting that increased halo spin could hinder gas accretion and slow star formation processes. However, this proposed scenario requires further investigation.

HI surveys conducted with single-dish telescopes, such as the Arecibo Legacy Fast Alfa Survey \citep[ALFALFA;][]{Giovanelli05,Haynes18} and the undergoing FAST All Sky HI survey \citep[FASHI;][]{Zhu23}, provide valuable HI spectra from numerous star-forming galaxies. These surveys offer crucial dynamical information on galaxies, facilitating the estimation of spin parameters for galaxies with varying star formation rates and enabling the study of halo spin's impact on star formation rate.

In this study, we utilize a semi-analytic approach to estimate halo spin for each HI-bearing galaxy cataloged in ALFALFA and investigate the relationship between halo spins and stellar densities of galaxies. Section~\ref{sec:2} introduces the sample data and outlines the methodology for estimating halo spin. Section~\ref{sec:3} presents a statistical analysis of the dependence of galaxy stellar densities on halo spins. Our findings are summarized in section~\ref{sec:4}.

\section{Data}
\label{sec:2}

\subsection{Sample and specific star formation rate}

We draw our galaxy sample from ALFALFA, a comprehensive HI survey covering 6,600 deg$^2$ at high Galactic latitudes. The ALFALFA \citep[$\alpha.$100;][]{Haynes18} catalog, released by \cite{Haynes18}, includes $\sim$31,500 sources with radial velocities below 18,000 km s$^{-1}$. For each source, the catalog provides the HI spectrum signal-to-noise ratio (SNR), cosmological distance, 50\% peak width of the HI line ($W_{50}$) corrected for instrumental effects, HI mass ($M_{\rm{HI}}$), among other properties.

\begin{figure}
   \centering
   \includegraphics[width=0.8\textwidth, angle=0]{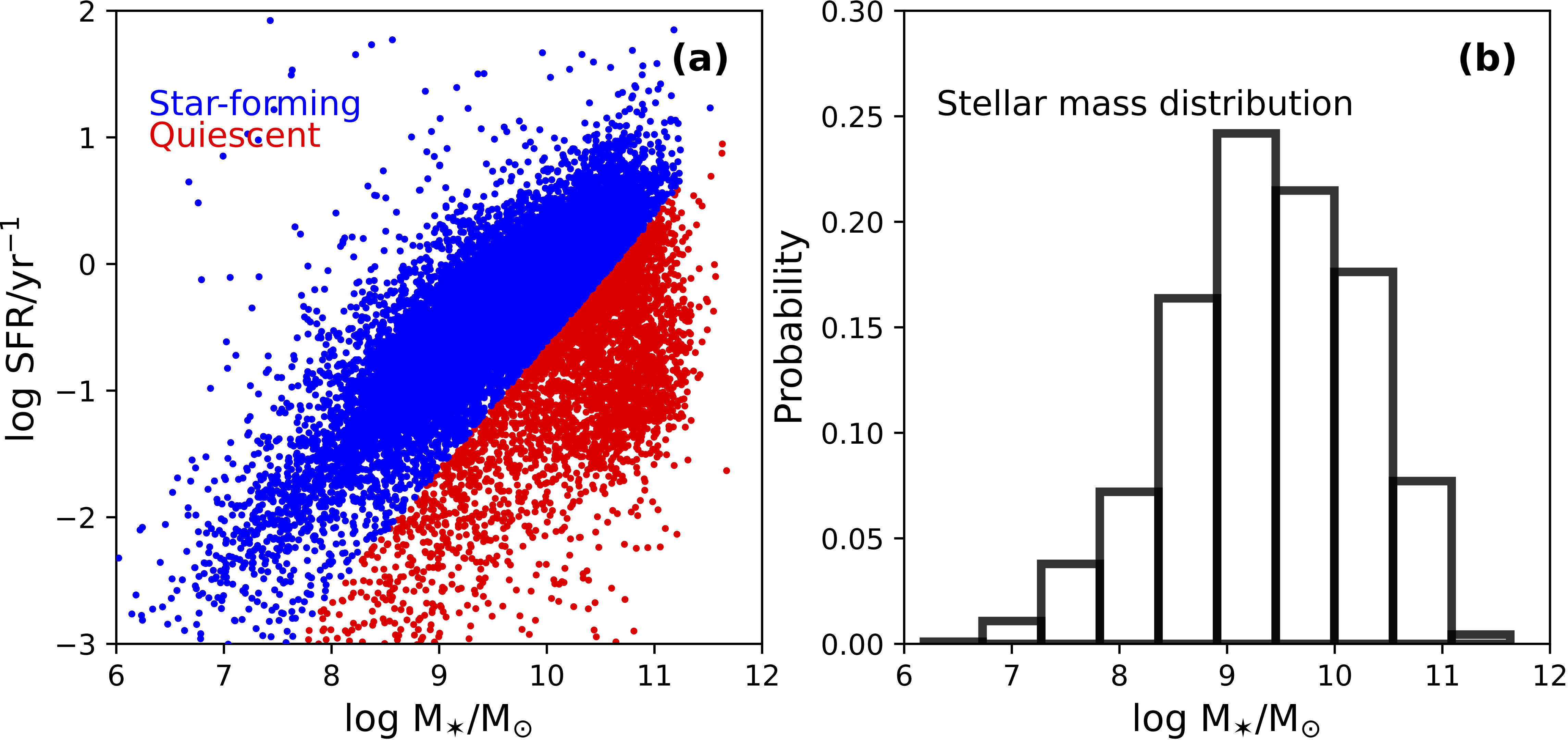}
   \caption{a). SFR vs. stellar mass for the star-forming (blue) and quiescent (red) galaxies. b). Stellar mass distribution for the star-forming galaxies in this study.}
   \label{stm_dis}
   \end{figure}

We match ALFALFA galaxies with MPA-JHU DR7 SDSS measurements to obtain star formation rates (SFRs) \citep{Brinchmann04} and stellar mass ($M_{\star}$) based on photometric fits. The specific SFR (sSFR) for each galaxy is then calculated as sSFR = SFR/$M_{\star}\ \rm(yr^{-1})$. To focus on star-forming galaxies, we select those with log sSFR $> \log(1/3t_{H(z=0)}) \approx -10.62$ \citep{Jing21}, where $t_{H(z=0)}$ represents the Hubble time at redshift 0. After applying these selection criteria, our sample comprises 15,787 star-forming galaxies, consisting of 8,187 low-mass ($M_{\star}<10^{9.5}\ M_{\odot}$) systems and 7,600 high-mass ($M_{\star}>10^{9.5}\ M_{\odot}$) counterparts. As illustrated in panel~a of Fig.~\ref{stm_dis}, the vast majority of these galaxies lie on the star-forming main sequence \citep{Noeske07,Schreiber15}, confirming their actively star-forming nature. Panel~b of Fig.~\ref{stm_dis} presents the stellar mass distribution of the selected sample.

\subsection{Rotation velocity and halo spin}

The rotation velocity is given by $V_{\rm{rot}} = W_{50}/2/\sin\phi$, with inclination $\phi$ of the HI disk estimated via the optical axis ratio $b/a$ (from \citealt{Durbala20}) and $\sin\phi = \sqrt{(1-(b/a)^2)/(1-q_0^2)}$, setting $\phi=90^{\circ}$ for $b/a \leq q_0$. We assume $q_0 \sim 0.2$ for massive galaxies and $q_0 \sim 0.4$ \citep{Rong24} for low-mass galaxies ($M_{\star}<10^{9.5} M_{\odot}$). We exclude galaxies with $\phi<50^{\circ}$ or low SNR ($<10$) to ensure accurate $V_{\rm{rot}}$ measurements.

Some HI-bearing galaxies exhibit velocity dispersion-dominated kinematics. These galaxies, identified by their HI line profiles exhibiting a `single-horned' shape \citep{ElBadry18}, pose challenges in accurately estimating rotation velocities and, consequently, halo spins. Following \citealt{Hua24}, we employ the kurtosis parameter $k_4<-1.0$ to restrict our analysis to robust subsets of galaxies with double-horned HI profiles to exclude potential contamination from dispersion-dominated systems. After applying all selection criteria, our final sample comprises 2,957 star-forming galaxies.

Assuming an isothermal halo model with negligible baryonic gravitational influence, the halo spin parameter $\lambda_{\rm{h}}$ is estimated as \citep{Hernandez07}:
\begin{equation}
\lambda_{\rm{h}}\simeq 21.8 \frac{R_{\rm{HI,d}}/{\rm kpc}}{(V_{\rm{rot}}/{\rm km s^{-1}})^{3/2}},
	\label{sam_HI} \end{equation}
where $R_{\rm{HI,d}}$ is the HI disk scale length, derived from:
\begin{equation}
	\Sigma_{\rm{HI}}(R)=\Sigma_{{\rm{HI}},0} {\rm{exp}}(-R/R_{{\rm{HI,d}}}),
\end{equation}
where $\Sigma_{{\rm{HI}},0}$ is the central surface density of the HI disk. The total HI mass $M_{\rm{HI}}$ is linked to the scale length as
\begin{equation} M_{\rm{HI}} = 2 \pi \Sigma_{{\rm{HI}},0} R_{{\rm{HI,d}}}^2 \label{HIeq_mass}. \end{equation} 
Additionally, we introduce the HI radius $r_{\rm{HI}}$, defined as the radius at which the HI surface density reaches $1\ \rm M_{\odot}\rm{pc^{-2}}$. $r_{\rm{HI}}$ is calculated using the observed $r_{\rm{HI}}$-$M_{\rm{HI}}$ relation \citep{Wang16, Gault21}: $\log r_{\rm{HI}}=0.51\log M_{\rm{HI}}-3.59$ \citep{Wang16,Gault21}. Therefore, at $r_{\rm{HI}}$, we have 
\begin{equation} \Sigma_{{\rm{HI}},0} {\rm{exp}}(-r_{\rm{HI}}/R_{{\rm{HI,d}}})=1\ \rm  M_{\odot}\rm{pc^{-2}}. \label{HIeq_3} \end{equation} 
By using equations~(\ref{HIeq_mass}) and (\ref{HIeq_3}), we can compute the value of $R_{{\rm{HI,d}}}$ for each galaxy in our sample, thereby enabling the estimation of the halo spin.


\begin{figure}
   \centering
   \includegraphics[width=0.8\textwidth, angle=0]{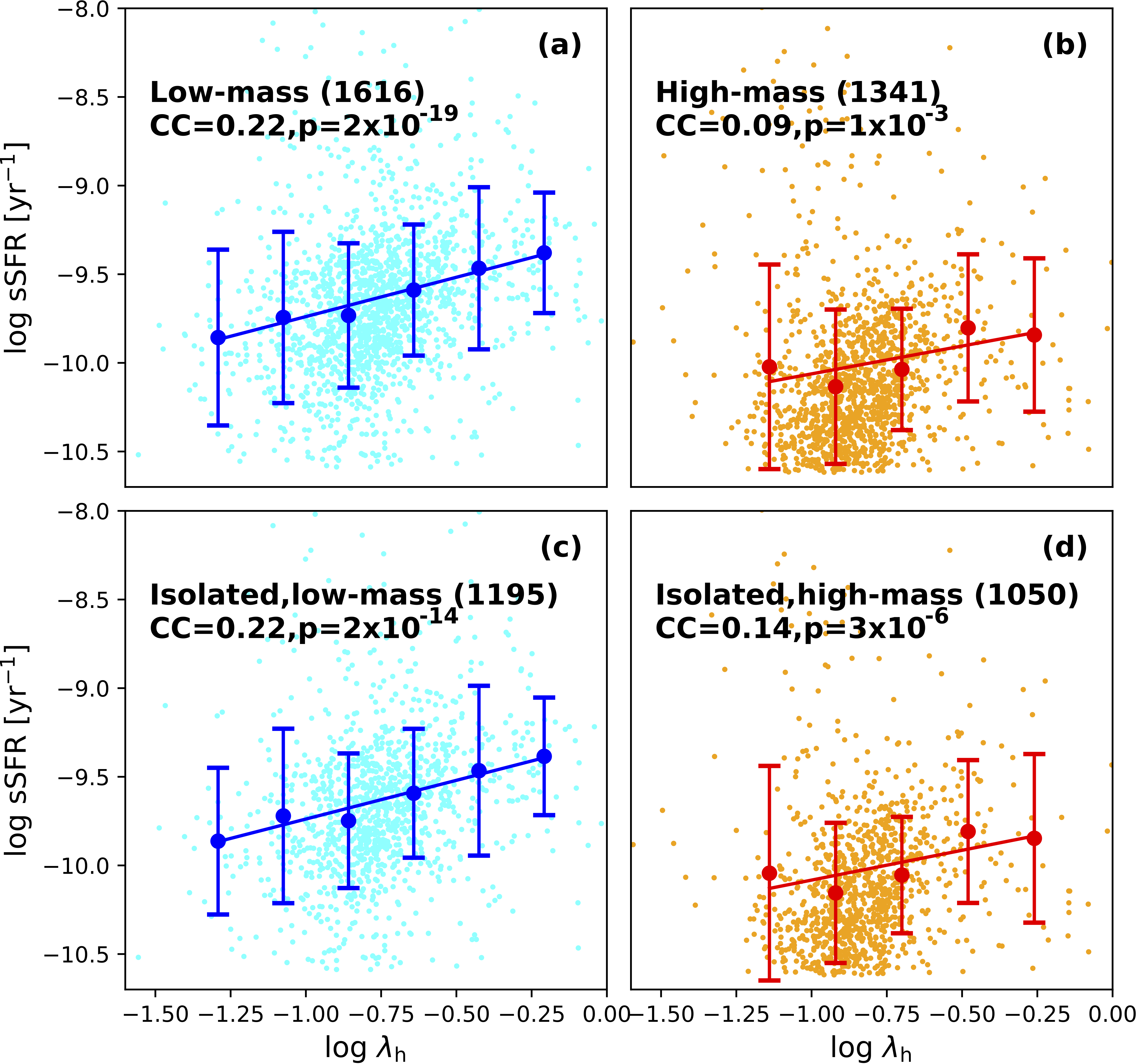}
   \caption{sSFR versus halo spin parameter for low-mass (left) and high-mass (right) galaxies. The top panels show the full star-forming galaxy sample, while the bottom panels focus on isolated galaxies. The numbers of galaxies are shown in the brackets of the corresponding panels. Median sSFR values with $1\sigma$ error bars for each bin in $\log \lambda_{\rm{h}}$ are represented. Best-fit linear trends are indicated by the corresponding lines. CC and $p$-value for the subsample are also shown in the corresponding panels.}
   \label{fig1}
   \end{figure}

   \begin{figure}
   \centering
   \includegraphics[width=0.8\textwidth, angle=0]{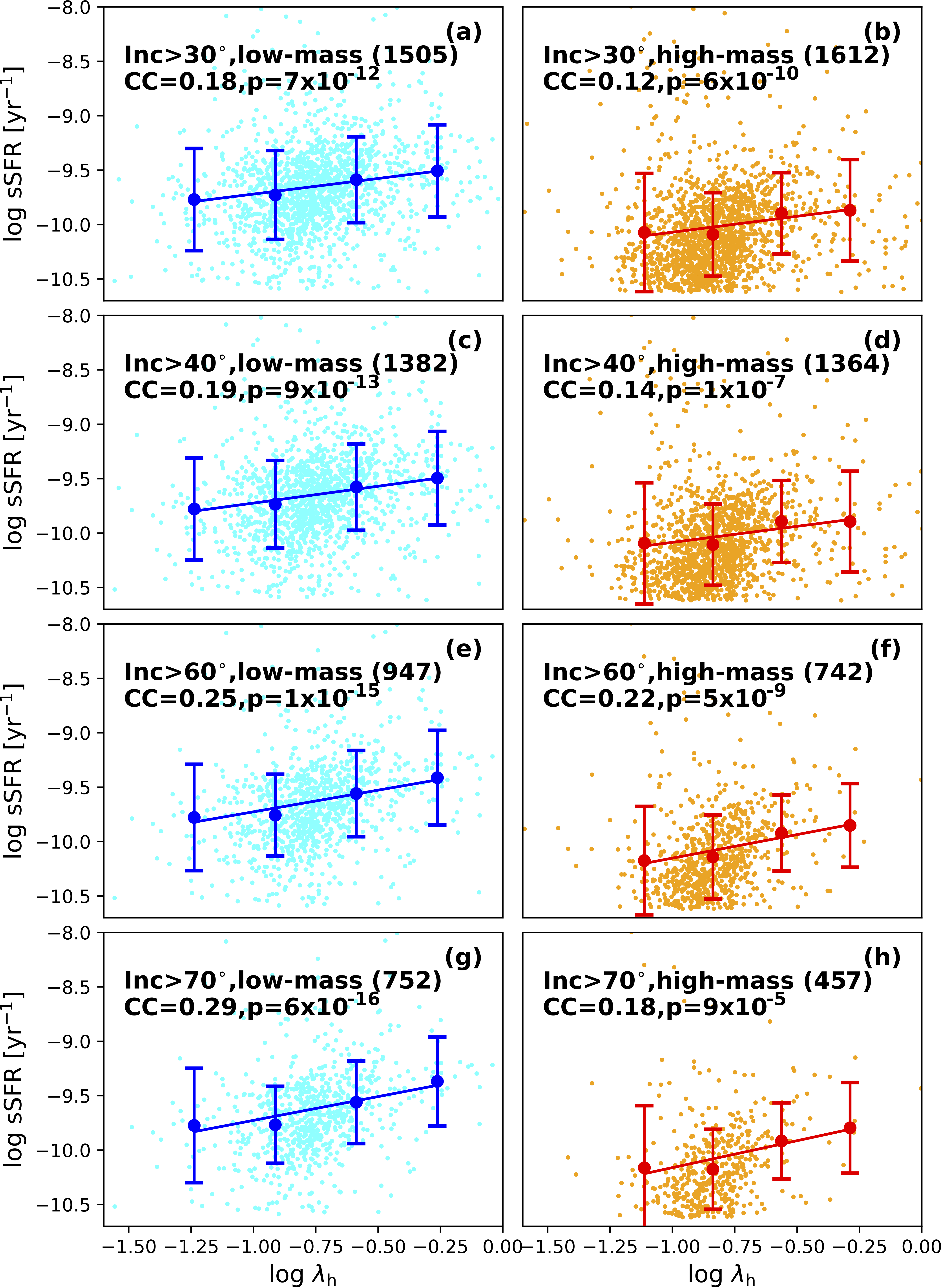}
   \caption{The correlations between sSFR and halo spin for low-mass (left) and high-mass (right) galaxies at the different inclination cutting thresholds. From the top panels to the bottom panels, we show the results for the isolated star-forming galaxies with the inclination cutting thresholds of $30^{\circ}$, $40^{\circ}$, $60^{\circ}$, $70^{\circ}$, respectively. Analogous to Fig.~\ref{fig1}, the median sSFR values with $1\sigma$ error bars for each bin in $\log \lambda_{\rm{h}}$ are represented, with the lines indicating the best-fit linear trends. CC and $p$-value for the subsample are also shown in the corresponding panels.}
   \label{test_inc}
   \end{figure}

   \begin{figure}
   \centering
   \includegraphics[width=0.8\textwidth, angle=0]{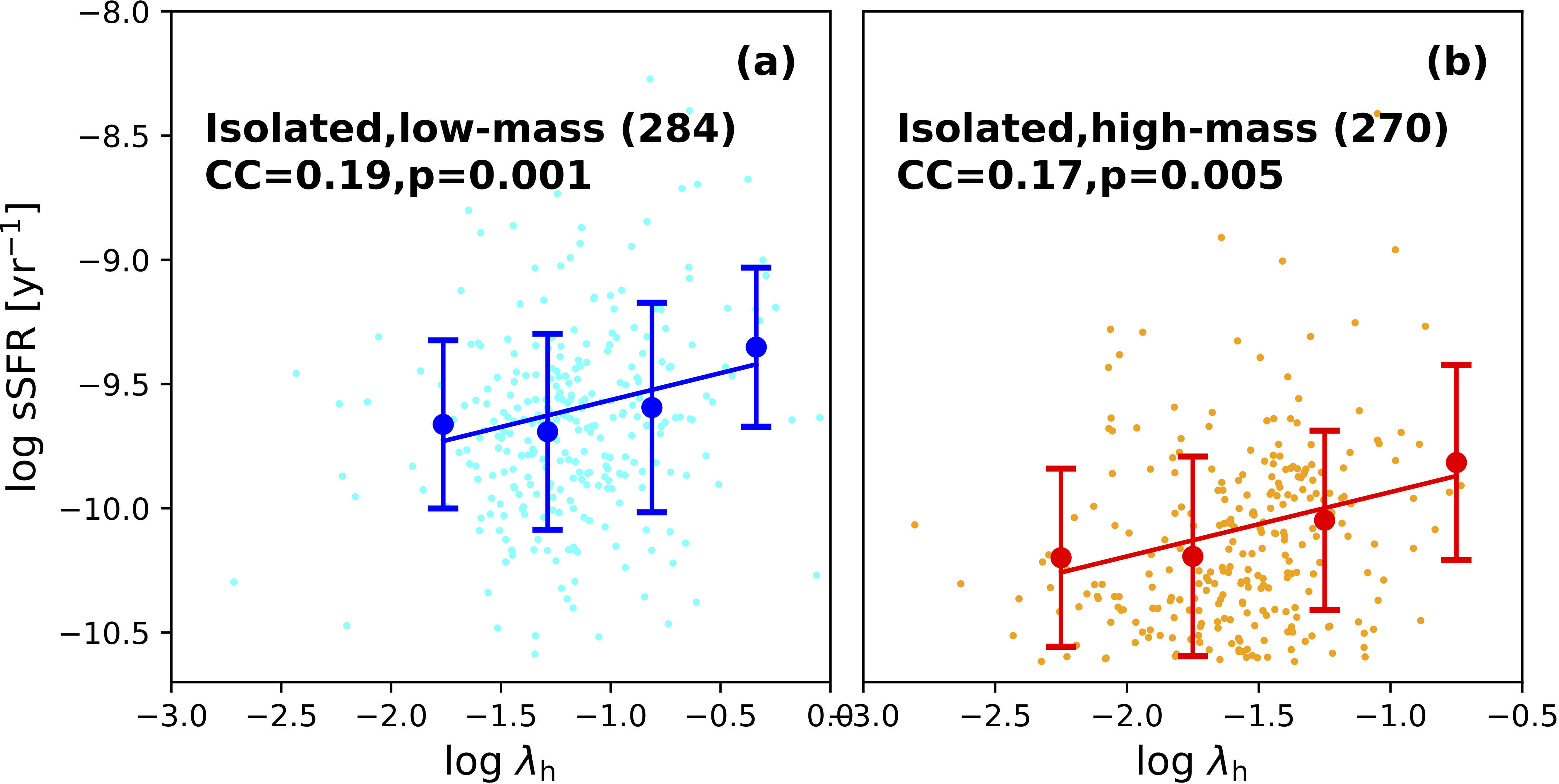}
   \caption{sSFR versus halo spin parameter calculated with the scale length of stellar disk, for isolated low-mass (a) and high-mass (b) star-forming galaxies.}
   \label{test_re}
   \end{figure}

\section{Results}
\label{sec:3}

Panels~a and b of Fig.~\ref{fig1} illustrate the sSFR-halo spin relation for low-mass ($M_{\star}<10^{9.5} M_{\odot}$) and high-mass ($M_{\star}>10^{9.5} M_{\odot}$) galaxies. Correlation analysis demonstrates a statistically significant, albeit weak, dependence of sSFR on halo spin, with correlation coefficients of CC$\approx 0.2$ and CC$\approx 0.1$ for the low-mass and high-mass regimes, respectively. The robustness of this trend is underscored by low $p$-values ($p=2\times10^{-19}$ and $p=1\times 10^{-3}$ for the low- and high-mass subsample, respectively), confirming that the correlation, while modest in strength, is highly significant in both regimes. Notably, the relationship seems to be more pronounced in low-mass systems, plausibly suggesting a stronger coupling between halo spin and star formation efficiency in dwarf galaxies compared to their more massive counterparts.

Since environment also affects galaxy properties \citep[e.g.,][]{Moore96, Mayer01, Mastropietro05, Smith15, Kazantzidis11, Hua25}, we control for environmental influences by using the galaxy group catalog by \citet{Saulder16}, applying a friends-of-friends algorithm to SDSS DR12 \citep{Alam15} and 2MASS Redshift Survey data \citep{Huchra12}, adjusted for biases (e.g., Malmquist bias). The analysis of 2,245 isolated galaxies-defined as systems located beyond three virial radii from any galaxy group-reveals a consistent, albeit weak, correlation between sSFR and halo spin (Fig.~\ref{fig1}, panels~c \& d). Statistically significant trends are observed for both low-mass (CC$=0.22$, $p=2\times10^{-14}$) and high-mass (CC$=0.14$, $p=3\times10^{-6}$).

   \begin{figure}
   \centering
   \includegraphics[width=0.8\textwidth, angle=0]{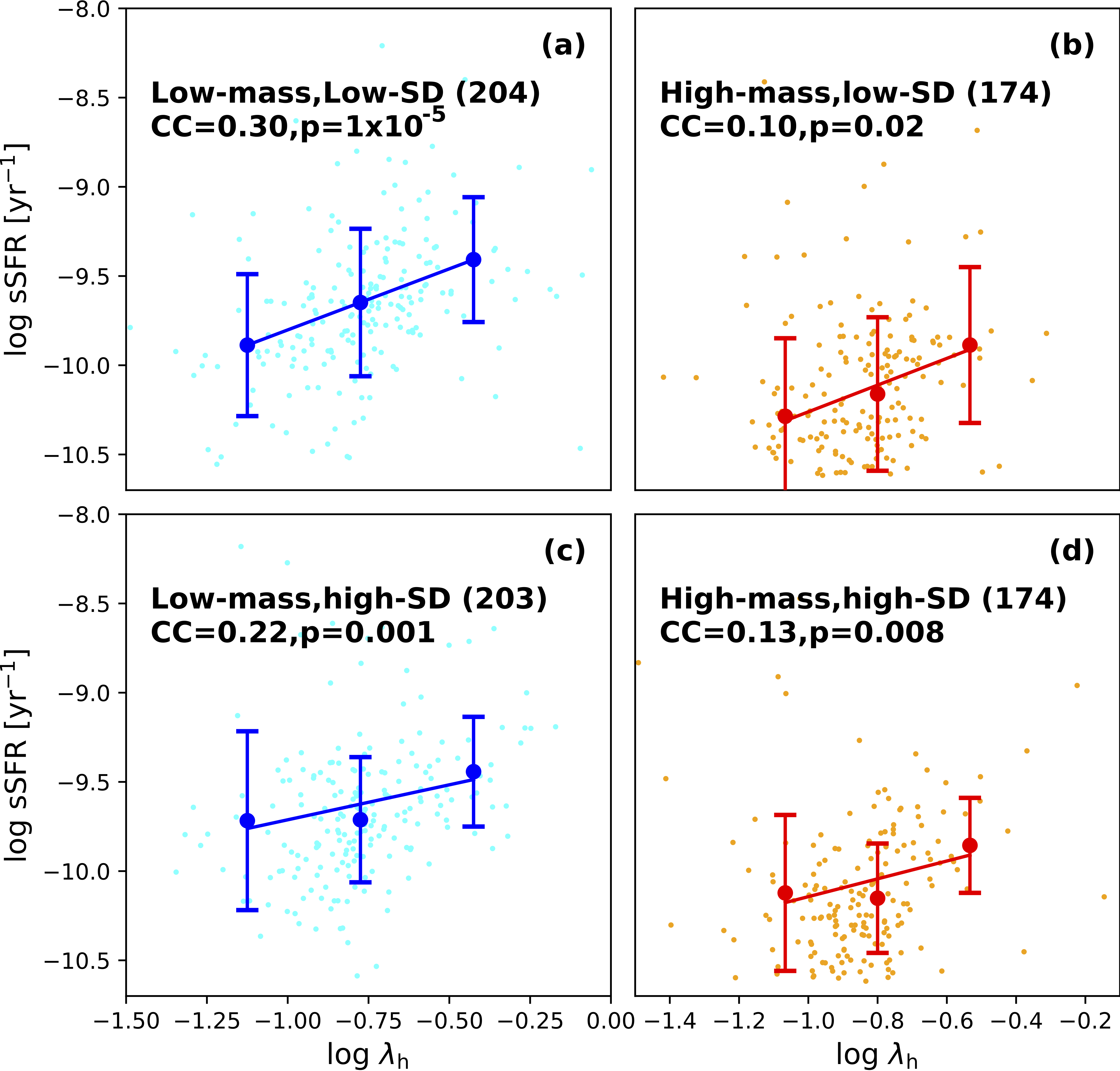}
   \caption{sSFR versus halo spin parameter for low-mass (left) and high-mass (right) star-forming galaxies. The upper and lower panels show the correlation for the low and high surface density subsamples, respectively.}
   \label{test_ss}
   \end{figure}

   \begin{figure}
   \centering
   \includegraphics[width=0.8\textwidth, angle=0]{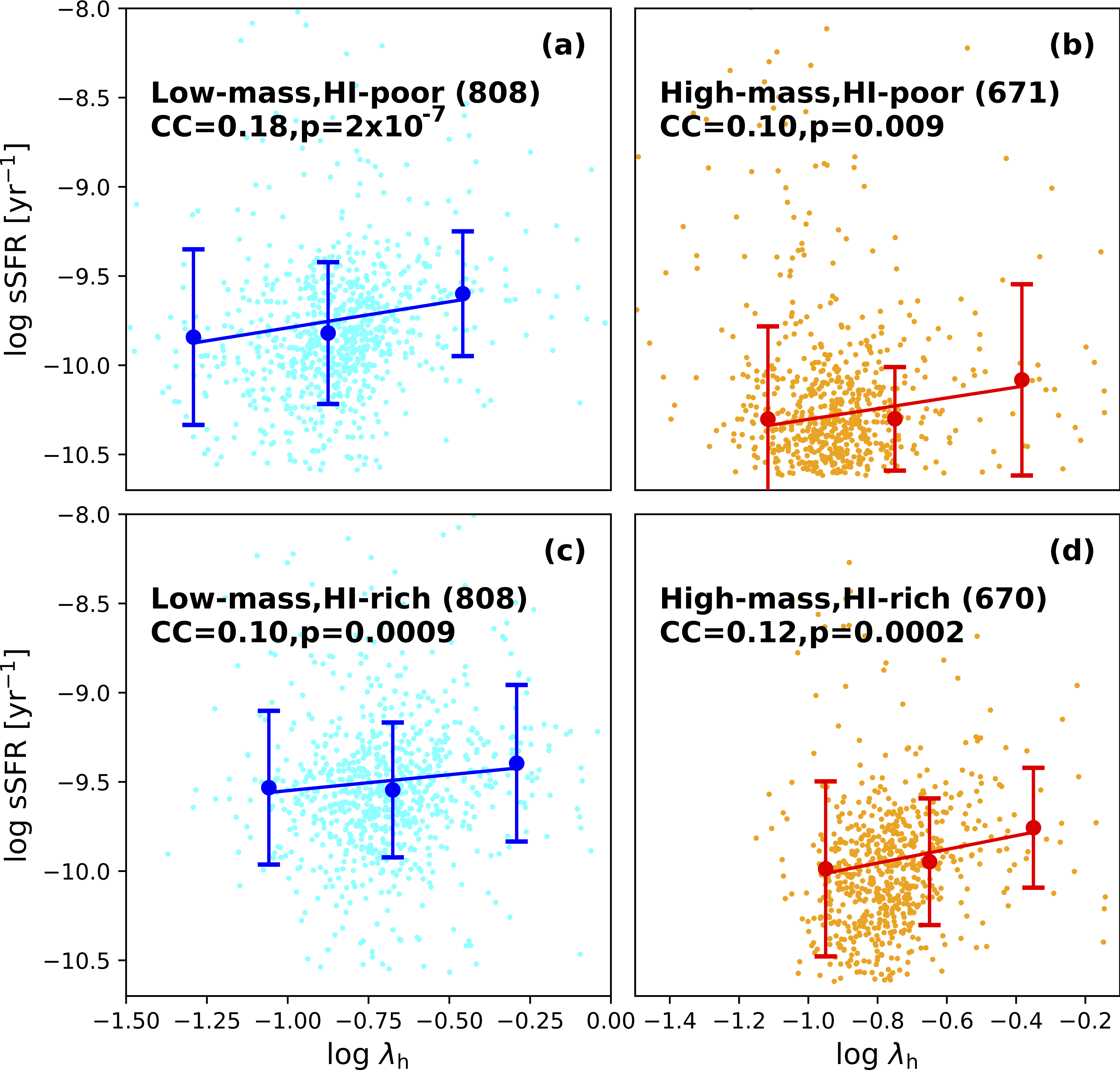}
   \caption{Analogous to Fig.~\ref{test_ss}, this plot shows the sSFR versus halo spin parameter for low-mass (left) and high-mass (right) star-forming galaxies. The upper and lower panels show the correlation for the HI-poor and HI-rich subsamples, respectively.}
   \label{test_hif}
   \end{figure}

    \begin{figure}
   \centering
   \includegraphics[width=0.8\textwidth, angle=0]{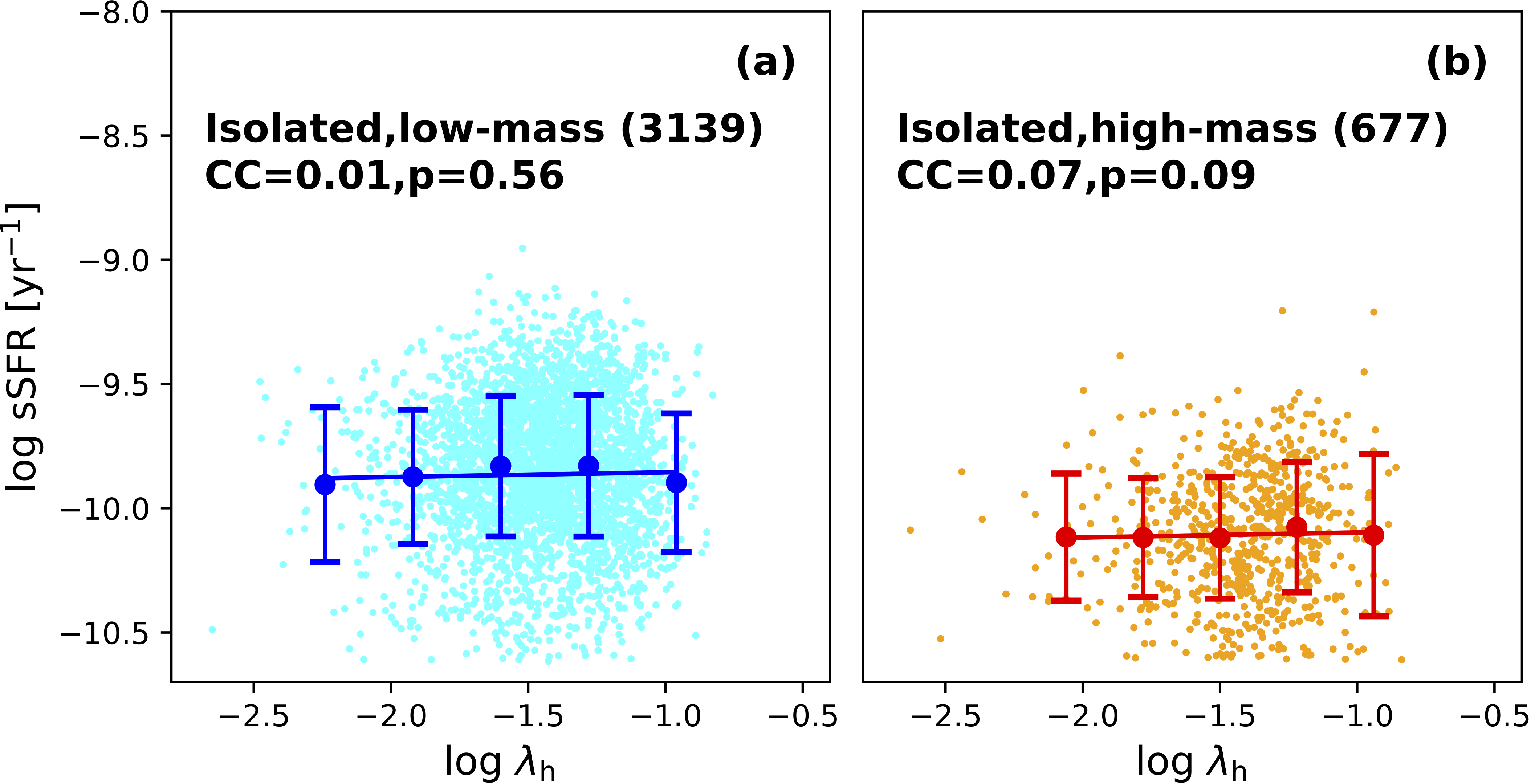}
   \caption{sSFR versus halo spin parameter for low-mass (left) and high-mass (right) isolated star-forming galaxies in TNG50 simulations.}
   \label{test_tng}
   \end{figure}

\section{Discussion and conclusion}
\label{sec:4}

Using ALFALFA HI data, we examine sSFR-halo spin relationships for a range of star-forming galaxies. Our analysis reveals a weak but statistically significant correlation between them, ndependent of environment. Notably, star-forming main sequence galaxies exhibit increasing sSFR with higher spin parameters, with this trend being more pronounced in low-mass systems compared to their high-mass counterparts.

To ensure the robustness of this finding, we conducted multiple validation tests. First, we examine the sSFR-spin correlation across multiple inclination thresholds ($30^{\circ}$, $40^{\circ}$, $60^{\circ}$, and $70^{\circ}$). As shown in Fig.~\ref{test_inc}, the statistically significant correlation persists at all tested inclination angles, despite showing relatively weak correlation coefficients (0.1$<$CC$<$0.3). This demonstrates that our results are robust against potential inclination-related biases in the HI kinematic measurements. Second, to verify the methodology dependence, we also recompute the spin parameter using stellar disk scale lengths ($R_{\star,d}$) instead of HI scale lengths ($R_{\rm{HI,d}}$) with equation~(\ref{sam_HI}). The stellar scale lengths are derived from effective radii ($R_{\rm{e}}$) following $R_{\star,d}\simeq R_{\rm{e}}/1.678$ \citep{Graham05}, where $R_{\rm{e}}$ values are taken from our previous work  \citep{Rong25}. While this alternative approach maintains the sSFR-spin correlation (Fig.~\ref{test_re}), the reduced sample size (due to limited $R_{\rm{e}}$ measurements for many ALFALFA galaxies) results in larger $p$-value and weaker statistical significance.

Third, given the established correlation between surface brightness/density and halo spin \citep[e.g.,][]{Cortese22,Rong25}, we must consider whether the observed sSFR-spin correlation could be artificially induced by sample incompleteness at the high-spin end—where galaxies with low surface brightness/density may be underrepresented \citep{Rong25}. To test this possibility, we split the star-forming galaxy sample into two equal-size subsamples based on stellar surface density (SD) using measurements from \cite{Rong25}, and examine the sSFR-spin relationship independently for both low-SD and high-SD subsamples. As illustrated in Fig.~\ref{test_ss}, both subsamples retain a significant correlation between sSFR and $\lambda_{\rm h}$, demonstrating that the observed trend is intrinsic rather than an artifact of observational biases.

Finally, building upon the established correlation between HI-to-stellar mass ratio and halo spin \citep{Liu25}, further test the intrinsic nature of the sSFR-spin relationship by dividing our isolated star-forming sample into HI-rich and HI-poor subsamples of equal size. Crucially, both subsamples maintain a statistically significant correlation between sSFR and $\lambda_{\rm h}$ (Fig.~\ref{test_hif}). This robust confirmation suggests that the sSFR-spin connection reflects fundamental galactic physics rather than secondary dependencies on HI mass fraction.

Our results show excellent agreement with previous findings from MaNGA IFU data \citep{Cortese22}, confirming the existence of an sSFR-spin correlation in observational data. However, this relationship is notably absent in the Illustris-TNG50 cosmological simulation  \citep{Nelson19}, where we find no significant correlation between sSFR and $\lambda_{\rm{h}}$, where we find no significant correlation between sSFR and $\lambda_{\rm{h}}$ for isolated star-forming galaxies at $z\sim 0$. Fig.~\ref{test_tng}) shows the sSFRs (calculated within 2$R_{\rm{e}}$) versus $\lambda_{\rm{h}}$ \citep[estimated with the same method of][]{Rodriguez-Gomez22} for galaxies. The weak correlation strength (CC$\sim 0$) and high $p$-values suggest that current simulations may not fully capture the complex baryonic physics governing this relationship, particularly in their treatment of star formation thresholds and feedback mechanisms.

The observed positive correlation between halo spin and sSFR may be understood through several interconnected physical mechanisms. Galaxies residing in high-spin halos naturally develop more extended, dynamically colder gaseous disks \citep{Mo98,Teklu15}. These morphological characteristics plausibly promote two key conditions for enhanced star formation: (1) elevated gas surface densities that exceed local star formation thresholds, and (2) prolonged gas reservoirs that sustain star formation over extended timescales \citep{Kennicutt98,Krumholz05}. Recent studies \citep{Rong25,Liu25} provide empirical support for this scenario, demonstrating that high-spin halos indeed host more extended disks with greater gas fractions. The stronger spin-sSFR correlation in low-mass galaxies likely reflects their dominant cold-mode accretion \citep{Keres05}. In these systems, the star formation efficiency becomes especially sensitive to angular momentum transport and disk stability \citep{Dekel06} as their shallower potential wells are less effective at thermalizing incoming gas flows.

While our analysis demonstrates a clear sSFR-spin correlation in typical star-forming galaxies, we note an important exception: isolated ultra-diffuse galaxies: ultra-diffuse galaxies (UDGs). Although UDGs are theoretically predicted to possess high halo spins \citep{Rong17a}, they consistently exhibit suppressed star formation rates \citep{Rong20}.  This apparent contradiction can be reconciled by considering that UDGs represent a rare population \citep{vanDokkum15} with distinct evolutionary pathways, and that our sample selection criteria explicitly excluded low-sSFR systems, including UDGs. It is worth to note that the sSFR-spin correlation appears to be valid only for actively star-forming galaxies.


\begin{acknowledgements}
Y.R. acknowledges supports from the NSFC grants 12273037 and 12522302, the CAS Pioneer Hundred Talents Program (Category B), the USTC Research Funds of the Double First-Class Initiative. This work is supported by the China Manned Space Program with grant no. CMS-CSST-2025-A06 and CMS-CSST-2025-A08.
\end{acknowledgements}


\begin{thebibliography}{99}

 

\bibitem[\protect\citeauthoryear{Alam et al.}{2015}]{Alam15} {{Alam}, M. P.} \emph{et~al.} 2015, \apjs, 219, 12

\bibitem[\protect\citeauthoryear{Amorisco \& Loeb}{2016}]{Amorisco16} {{Amorisco}, N. C. \& {Loeb}, A.} 2016, \mnras, 459, L51


\bibitem[\protect\citeauthoryear{Benavides et al.}{2023}]{Benavides23} {{Benavides}, J. A., {Sales}, L. V., {Abadi}, M. G., {Marinacci}, F., {Vogelsberger}, M., {Hernquist}, L.} 2023, \mnras, 522, 1033

\bibitem[\protect\citeauthoryear{Brinchmann et al.}{2004}]{Brinchmann04} Brinchmann, J., Charlot, S., White, S. D. M., Tremonti, C., Kauffmann, G., Heckman, T., Brinkmann, J. 2004, \mnras, 351, 1151

\bibitem[\protect\citeauthoryear{Cortese et al.}{2022}]{Cortese22} Cortese, L., Fraser-McKelvie, A., Woo, J., et al. 2022, MNRAS, 513, 3709

\bibitem[\protect\citeauthoryear{Dahlem et al.}{2006}]{Dahlem06} Dahlem, M., Lisenfeld, U., Rossa, J. 2006, A\&A, 457, 121

\bibitem[\protect\citeauthoryear{Dekel et al.}{2006}]{Dekel06} Dekel, A., Birnboim, Y. 2006, MNRAS, 368, 2

\bibitem[\protect\citeauthoryear{Desmond et al.}{2017}]{Desmond17} Desmond, H., Mao Y.-Y., Wechsler R. H., Crain R. A., Schaye J. 2017,
\mnras, 471, L11

\bibitem[\protect\citeauthoryear{Di Cintio et al.}{2019}]{DiCintio19} Di Cintio, A., Brook, C. B., Macci\`o, A. V., Dutton, A. V., Cardona-Barrero, S. 2019, MNRAS, 486, 2535

\bibitem[\protect\citeauthoryear{Diemand et al.}{2005}]{Diemand05} Diemand, J., Madau, P., Moore, B. 2005, \mnras, 364, 367

\bibitem[\protect\citeauthoryear{Durbala et al.}{2020}]{Durbala20} {{Durbala}, A., {Finn}, R. A., {Crone Odekon}, M., {Haynes}, M. P., {Koopmann}, R. A., {O'Donoghue}, A. A.} 2020, \aj, 160, 271

\bibitem[\protect\citeauthoryear{El-Badry et al.}{2018}]{ElBadry18} {{El-Badry}, K.} \emph{et~al.} 2018, \mnras, 473, 1930

\bibitem[\protect\citeauthoryear{Gault et al.}{2021}]{Gault21} {{Gault}, L.} \emph{et~al.} 2021, \aj, 909, 19

\bibitem[\protect\citeauthoryear{Giovanelli et al.}{2005}]{Giovanelli05} {{Giovanelli}, R.} \emph{et~al.} 2005, \aj, 130, 6

 \bibitem[\protect\citeauthoryear{Gonz\'alez \& Padilla}{2009}]{Gonzalez09} Gonz\'alez, R. E., Padilla, N. D. 2009, \mnras, 397, 1498

 \bibitem[\protect\citeauthoryear{Graham \& Driver}{2005}]{Graham05} Graham, A. W., Driver, S. P. 2005, Publications of the Astronomical Society of Australia, 22, 118

\bibitem[\protect\citeauthoryear{Grudi\'c et al.}{2018}]{Grudic18} Grudi\'c, M. Y., Hopkins, P. F., Faucher-Gigu\`ere, C.-A., Quataert, E., Murray, N., Kere\v{s}, D. 2018, MNRAS, 475, 3511

\bibitem[\protect\citeauthoryear{{Guo} et al.}{2011}]{Guo11} {{Guo}, Q.}, \emph{et~al.} 2011, \mnras, 413, 101

\bibitem[\protect\citeauthoryear{Haynes et al.}{2018}]{Haynes18} {{Haynes}, M. P.} \emph{et~al.} 2018, \apj, 861, 49

\bibitem[\protect\citeauthoryear{Hayward et al.}{2011}]{Hayward11} Hayward, C. C., Kere\v{s}, D., Jonsson, P., Narayanan, D., Cox, T. J., Hernquist, L. 2011, ApJ, 743, 159

\bibitem[\protect\citeauthoryear{Hernandez et al.}{2007}]{Hernandez07} Hernandez, X., Park, C., Cervantes-Sodi, B., \& Choi, Y.-Y. 2007, \mnras, 375, 163

\bibitem[\protect\citeauthoryear{Hjorth et al.}{2014}]{Hjorth14} Hjorth, J., Gall, C., Michalowski, M. J. 2014, \apjl, 872, L23

\bibitem[\protect\citeauthoryear{Hopkins et al.}{2012}]{Hopkins12} Hopkins, P. F., Quataert, E., Murray, N. 2012, \mnras, 421, 3488

\bibitem[\protect\citeauthoryear{Hua et al.}{2025a}]{Hua24} Hua, Z., Rong, Y., Hu, H.-J. 2025a, MNRAS, 538, 775

\bibitem[\protect\citeauthoryear{Hua et al.}{2025b}]{Hua25} Hua, Z., Rong, Y., Hu, H.-J. 2025b, RAA, 25, 041001

\bibitem[\protect\citeauthoryear{Huchra et al.}{2012}]{Huchra12} {{Huchra}, J. P.} \emph{et~al.} 2012, \apjs, 199, 26

\bibitem[\protect\citeauthoryear{Hunter et al.}{2012}]{Hunter12} Hunter, D. A., \emph{et~al.} 2012, \aj, 144, 134

\bibitem[\protect\citeauthoryear{Jiang et al.}{2019}]{Jiang192} Jiang, F., et al. 2019, \mnras, 488, 4801

\bibitem[\protect\citeauthoryear{Jing et al.}{2021}]{Jing21} Jing, Y.-J., Rong, Y., Wang, J., Guo, Q., Gao, L. 2021, RAA, 21, 218

\bibitem[\protect\citeauthoryear{Kazantzidis et al.}{2011}]{Kazantzidis11} Kazantzidis, S., Lokas, E., Callegari, S., Mayer, L., Moustakas, L. 2011, \apj, 726, 98

\bibitem[\protect\citeauthoryear{Kennicutt}{1998}]{Kennicutt98} Kennicutt, R. C. Jr. 1998, ApJ, 498, 541

\bibitem[\protect\citeauthoryear{Kere\v{s} et al.}{2005}]{Keres05} Kere\v{s}, D., Katz, N., Weinberg, D. H., Dav\'e, R. 2005, MNRAS, 363, 2

\bibitem[\protect\citeauthoryear{Kim \& Lee}{2013}]{Kim13} Kim, J.-h. \& Lee, J. 2013, \mnras, 432, 1701

\bibitem[\protect\citeauthoryear{Kimm et al.}{2009}]{Kimm09} Kimm, T., et al. 2009, \mnras, 394, 1131

\bibitem[\protect\citeauthoryear{Kravtsov et al.}{2018}]{Kravtsov18} Kravtsov, A. V., Vikhlinin, A. A., 2018, Astronomy Letters, 44, 8

\bibitem[\protect\citeauthoryear{Krumholz \& McKee}{2005}]{Krumholz05} Krumholz, M. R., McKee, C. F. 2005, ApJ, 630, 250


\bibitem[\protect\citeauthoryear{Liu et al.}{2025}]{Liu25} Liu, S., Rong, Y., Hua, Z., Hu, H. 2025, RAA, 25, 081001


\bibitem[\protect\citeauthoryear{Mastropietro et al.}{2005}]{Mastropietro05} Mastropietro, C., Moore, B., Mayer, L., Debattista, V., Piffaretti, R., Stadel, J. 2005, \mnras, 364, 607

\bibitem[\protect\citeauthoryear{Mayer et al.}{2001}]{Mayer01} Mayer, L., Governato, F., Colpi, M., Moore, B., Quinn, T., Wadsley, J., Stadel, J., Lake, G. 2001, \apj, 547, L123

\bibitem[\protect\citeauthoryear{Mayer et al.}{2007}]{Mayer07} Mayer, L., Kazantzidis, S., Mastropietro, C., Wadsley, J. 2007, Nature, 445, 738


\bibitem[\protect\citeauthoryear{Mo et al.}{1998}]{Mo98} {{Mo}, H. J., {Mao}, S. D. \& {White}, S. D. M.} 1998, \mnras, 295, 319

\bibitem[\protect\citeauthoryear{Moore et al.}{1996}]{Moore96} Moore, B., Katz, N., Lake, G., et al. 1996, Nature, 379, 613 

 
\bibitem[\protect\citeauthoryear{Nelson et al.}{2019}]{Nelson19} Nelson, D., et al. 2019, Comput. Astrophys. Cosmol., 6, 2


\bibitem[\protect\citeauthoryear{Noeske et al.}{2007}]{Noeske07} Noeske, K. G., Weiner, B. J., Faber, S. M., et al. 2007, ApJL, 660, 43


\bibitem[\protect\citeauthoryear{Peng \& Renzini}{2020}]{Peng20} Peng, Y.-J., Renzini, A. 2020, \mnras, 491L, 51

\bibitem[\protect\citeauthoryear{Rodriguez-Gomez et al.}{2022}]{Rodriguez-Gomez22} Rodriguez-Gomez, V., et al. 2022, MNRAS, 512, 5978


\bibitem[\protect\citeauthoryear{{Rong} et al.}{2025}]{Rong25} Rong, Y., Hua, Z., Hu, H. 2025, RAA, 25, 011001

\bibitem[\protect\citeauthoryear{{Rong} et al.}{2017}]{Rong17a} {{Rong}, Y., {Guo}, Q., {Gao}, L., {Liao}, S., {Xie}, L., {Puzia}, T. H., {Sun}, S., {Pan}, J.} 2017, \mnras, 470, 4231


\bibitem[\protect\citeauthoryear{{Rong} et al.}{2020}]{Rong20} {Rong}, Y., {Zhu}, K., {Johnston}, E. J., {Zhang}, H.-X., {Cao}, T., {Puzia}, T. H., {Galaz}, G. 2020, ApJL, 899, L12


\bibitem[\protect\citeauthoryear{Rong et al.}{2024a}]{Rong24a} Rong, Y., Hu, H., He, M., Du, W., Guo, Q. Wang, H.-Y., Zhang, H.-X., Mo, H. 2024a, arXiv:2404.00555

\bibitem[\protect\citeauthoryear{Rong et al.}{2024b}]{Rong24} Rong, Y., He, M., Hu, H., Zhang, H.-X., Wang, H.-Y. 2024b, arXiv:2409.00944

\bibitem[\protect\citeauthoryear{Sales et al.}{2022}]{Sales22} Sales, L. V., Wetzel, A., Fattahi, A. 2022, Nature Astronomy, 6, 897

\bibitem[\protect\citeauthoryear{Saulder et al.}{2016}]{Saulder16} {{Saulder}, C., {van Kampen}, E., {Chilingarian}, I. V., {Mikske}, S., {Zeilinger}, W. W.} 2016, \aap, 596, A14

\bibitem[\protect\citeauthoryear{Sawala et al.}{2015}]{Sawala15} Sawala, T., et al. 2015, \mnras, 448, 2941

\bibitem[\protect\citeauthoryear{Schreiber et al.}{2015}]{Schreiber15} Schreiber, C., Elbaz, D., Pannella, M., et al. 2016, A\&A, 589, A35

\bibitem[\protect\citeauthoryear{Smith et al.}{2015}]{Smith15} Smith, R., S\'anchez-Janssen, R., Beasley, M. A., et al. 2015, \mnras, 454, 2502

\bibitem[\protect\citeauthoryear{Springel}{2000}]{Springel00} Springel, V. 2000, \mnras, 312, 859

\bibitem[\protect\citeauthoryear{Teklu et al.}{2015}]{Teklu15} Teklu, A. F., Remus, R.-S., Dolag, K., Beck, A. M., Burkert, A., Schmidt, A. S., Schulze, F., Steinborn, L. K. 2015, ApJ, 812, 29

\bibitem[\protect\citeauthoryear{Tinker et al.}{2017}]{Tinker17} Tinker, J. L., Wetzel, A. R., Conroy, C. 2017, \mnras, 472, 2504


\bibitem[\protect\citeauthoryear{van Dokkum}{2015}]{vanDokkum15} van Dokkum, P. G., Abraham, R., Merritt, A., Zhang, J., Geha, M., Conroy, C. 2015, ApJL, 798, 45

\bibitem[\protect\citeauthoryear{van den Bosch}{1998}]{vandenBosch98} van den Bosch, F. C. 1998, \apj, 507, 601

\bibitem[\protect\citeauthoryear{Wang et al.}{2016}]{Wang16} {{Wang}, J., {Koribalski}, B. S., {Serra}, P., {van der Hulst}, T., {Roychowdhury}, S., {Kamphuis}, P., {Chengalur}, J. N.} 2016, \mnras, 460, 2143

\bibitem[\protect\citeauthoryear{Yang et al.}{2023}]{Yang23} Yang, H., Gao, L., Frenk, C. S., Grand, R. J. J., Guo, Q., Liao, S., Shao, S. 2023, \mnras, 518, 5253

\bibitem[\protect\citeauthoryear{Zhang et al.}{2024}]{Zhu23} Zhang, C.-P., Zhu, M., Jiang, P., et al. 2024, Science China Physics, Mechanics \& Astronomy, 67, 219511

\end{thebibliography}

\label{lastpage}

\end{document}